\documentclass[a4paper,11pt]{article}
\usepackage{pos}
\usepackage{bibspacing}

\title{Heavy-flavored emissions in hybrid collinear/high-energy factorization}

\author[a,b]{A. D. Bolognino}
\author[c,d,e]{F. G. Celiberto}
\author*[a,b]{M. Fucilla}
\author[f]{D. Yu. Ivanov}
\author[a,b]{A. Papa}

\affiliation[a]{Dipartimento di Fisica, Università della Calabria,
I-87036 Arcavacata di Rende, Cosenza, Italy}

\affiliation[b]{Istituto Nazionale di Fisica Nucleare, Gruppo collegato di Cosenza,
I-87036 Arcavacata di Rende, Cosenza, Italy}

\affiliation[c]{European Centre for Theoretical Studies in Nuclear Physics and Related Areas (ECT*),
I-38123 Villazzano, Trento, Italy}

\affiliation[d]{Fondazione Bruno Kessler (FBK), I-38123 Povo, Trento, Italy}

\affiliation[e]{INFN-TIFPA Trento Institute of Fundamental Physics and Applications,
I-38123 Povo, Trento, Italy}

\affiliation[e]{Sobolev Institute of Mathematics, 630090 Novosibirsk, Russia}

\emailAdd{ad.bolognino@unical.it}
\emailAdd{francescogiovanni.celiberto@unipv.it}
\emailAdd{michael.fucilla@unical.it}
\emailAdd{d-ivanov@math.nsc.ru}
\emailAdd{alessandro.papa@fis.unical.it}

\abstract{Heavy-flavored emissions have been always considered as an excellent channel to test properties of Quantum chromodynamics (QCD) at present and future colliders. Among different regimes, in which heavy-flavor production can be investigated, we focus our attention on the semi-hard one, where $s \gg Q^2 \gg \Lambda_{{\rm{QCD}}}$ ($s$ is the squared center-of-mass energy, ${Q^2}$ a (set of) hard scale(s) characteristic of the process and $\Lambda_{{\rm{QCD}}}$ the QCD mass scale). Here, we build predictions in a hybrid collinear/high-energy factorization, in which the standard collinear description is supplemented by the Balitsky-Fadin-Kuraev-Lipatov resummation of large energy logarithms. The definition and the study of observables sensitive to high-energy dynamics in the context of heavy-flavor physics has the double advantage of (\emph{i}) allowing to get a stabilization of the BFKL series under higher-order corrections and (\emph{ii}) providing us with an auxiliary tool to investigate heavy-flavor production in wider kinematical ranges. Hence, we propose a scientific program on heavy-flavor physics at high energy with the goal of considering both open (heavy-jet) and bound states ($\Lambda$~baryons, heavy-light mesons and quarkonia).}

\FullConference{%
  *** The European Physical Society Conference on High Energy Physics (EPS-HEP2021), ***\\
  *** 26-30 July 2021 ***\\
  *** Online conference, jointly organized by Universität Hamburg and the research center DESY ***
}

\begin{document}
\maketitle

\vspace{-0.25cm}
\section{Introduction}
In the TeV-range, which is to be considered at modern colliders, hadronic reactions can be explored in new kinematic regimes. A particularly interesting one is the so-called \textit{semi-hard regime}, characterized by a center-of-mass energy, $\sqrt{s}$, much larger than the hard scales of the process, $\{ Q \}$, which are, in turn, much larger than the QCD mass scale, $\Lambda_{{\rm{QCD}}}$. Here, large logarithms of the energy enter the perturbative series with powers increasing with the perturbative order, thus systematically compensating the smallness of the coupling. Therefore, a resummation to all orders which takes into account the effects of these large logarithms is required. The consolidated tool for this resummation is the Balitsky-Fadin-Kuraev-Lipatov (BFKL) approach~\cite{Fadin:1975cb,Kuraev:1976ge,Kuraev:1977fs,Balitsky:1978ic}, which allows for the inclusion of large energy logarithms both in the leading logarithmic approximation ($(\alpha_s \ln s)^n$ terms are resummed), LLA, and in the next-to-leading logarithmic approximation ($(\alpha_s \ln s)^n$ and $\alpha_s (\alpha_s \ln s)^n$ terms are resummed), NLA. In this framework, the cross sections of processes take a peculiar factorized form, given by the convolution of two process-dependent impact factors, related to the transition of colliding particles into a precise final-state (in their fragmentation region), and a process-independent Green’s function.
A selection of reactions that can be considered at NLA level includes: 
the inclusive hadroproduction of two jets well separated in rapidity
(Mueller–Navelet channel~\cite{Mueller:1986ey}), for which several phenomenological analyses have appeared so
far~\cite{Colferai:2010wu,Caporale:2012ih,Ducloue:2013hia,Ducloue:2013bva,Caporale:2014gpa,Caporale:2015uva,Celiberto:2015yba,Celiberto:2015mpa,Celiberto:2016ygs,Celiberto:2016vva,Caporale:2018qnm}, the inclusive detection of two light-charged rapidity-separated hadrons~\cite{Celiberto:2016hae,Celiberto:2016zgb,Celiberto:2017ptm,Celiberto:2017ydk} or of a rapidity-separated pair formed by a light-charged hadron and a jet~\cite{Bolognino:2018oth,Bolognino:2019yqj,Bolognino:2019cac}, the inclusive
production of rapidity-separated $\Lambda$-$\Lambda$ or $\Lambda$-jet pairs~\cite{Celiberto:2020rxb}.
Settling for only a partial inclusion of next-to-leading effects, new channels open up, such as three- and four-jet hadroproduction~\cite{Caporale:2015vya,Caporale:2015int,Caporale:2016soq,Caporale:2016xku,Caporale:2016zkc}, $J/\Psi$-jet~\cite{Boussarie:2017oae},
Drell--Yan-jet~\cite{Golec-Biernat:2018kem}, Higgs-jet~\cite{Celiberto:2020tmb}, and heavy-quark pair production~\cite{Celiberto:2017nyx,Bolognino:2019ouc,Bolognino:2019yls,Bolognino:2021mrc}.
In these reactions two objects with a large separation in rapidity are inclusively tagged, together with an undetected hadronic system. They can be investigated via the so-called \textit{hybrid} collinear/high-energy factorization, where collinear ingredients, such as parton distribution functions (PDFs), fragmentation functions (FFs) and jet functions (JFs), enter the definiton of BFKL impact factors. Another class of reactions that can be studied in the BFKL approach are the so-called single-forward emissions, where the gluon content in the proton is accessed via the \emph{unintegrated gluon distribution} (UGD)~\cite{Bolognino:2018rhb,Celiberto:2018muu,Bolognino:2021niq}.
Below, we will focus on heavy flavored emissions, in particular to $\Lambda_c$-baryon productions. The reason for our interest lies in the fact that these reactions present a novel feature in BFKL phenomenology, allowing for a (partial) stabilization of the series, under inclusion of high-order corrections and scale variations. 

\vspace{-0.25cm}
\section{$\Lambda_c$-$\Lambda_c$ and $\Lambda_c$-jet production in VFNS: Theoretical set-up}

We considered two hadronic reactions:
\begin{equation}
\label{process_LL}
 {\rm proton}(P_a) + {\rm proton}(P_b) \to \Lambda_c^\pm(p_1, y_1) + X + \Lambda_c^\pm(p_2 , y_2) \;,
\end{equation}
\vspace{-0.75cm}
\begin{equation}
\label{process_LJ}
 {\rm proton}(P_a) + {\rm proton}(P_b) \to \Lambda_c^\pm(p_1, y_1) + X + {\rm jet}(p_2 , y_2) \;,
\end{equation}

We consider these channels in the high $p_T$-regime, which justifies the use of a \textit{zero-mass variable flavor number scheme} (ZM-VFNS), in which all five active quarks are considered as massless, and therefore the fragmentation in the $\Lambda_c$ occurs from light particles.
At variance with the standard collinear approach, in our treatment we start from a high-energy factorization which emerges inside the BFKL formalism and we then add collinear ingredients at the level of impact factors. In fact, the ``pure" BFKL treatment allows us to build partonic distributions, which are not infrared-safe quantities. We need to reabsorb divergences associated with initial and also final (since we are not totally inclusive) state radiation, at the level of impact factors, thus defining ``hadronic impact factors'', given by a convolution of BFKL partonic impact factors with PDFs and FFs.
It is convenient to write the cross section as a Fourier series of the azimuthal-angle coefficients, ${\cal C}_{n \ge 0}$
\begin{equation}
 \label{dsigma_Fourier}
 \frac{d \sigma}{d y_1 d y_2 d |\vec p_1| d |\vec p_2| d \varphi_1 d \varphi_2} =
 \frac{1}{(2\pi)^2} \left[{\cal C}_0 + 2 \sum_{n=1}^\infty \cos (n \varphi)\,
 {\cal C}_n \right]\, ,
\end{equation}
where $\varphi_{1,2}$ are the azimuthal angles of the tagged objects and $\varphi \equiv \varphi_1 - \varphi_2 - \pi$. The definition of ${\cal C}_n$ in the $\overline{\rm{MS}}$-scheme can be found in \cite{Celiberto:2021dzy}.
We remark that the description of $\Lambda_c$ particles in terms of light-hadron impact factors is adequate, provided that energy scales are much larger than the $\Lambda_c$ mass. This condition is guaranteed by the transverse-momentum ranges of our interest.

\vspace{-0.25cm}
\section{$\Lambda_c$ production: Phenomenology and stabilization effects}
\begin{figure}
\centering

   \includegraphics[scale=0.40,clip]{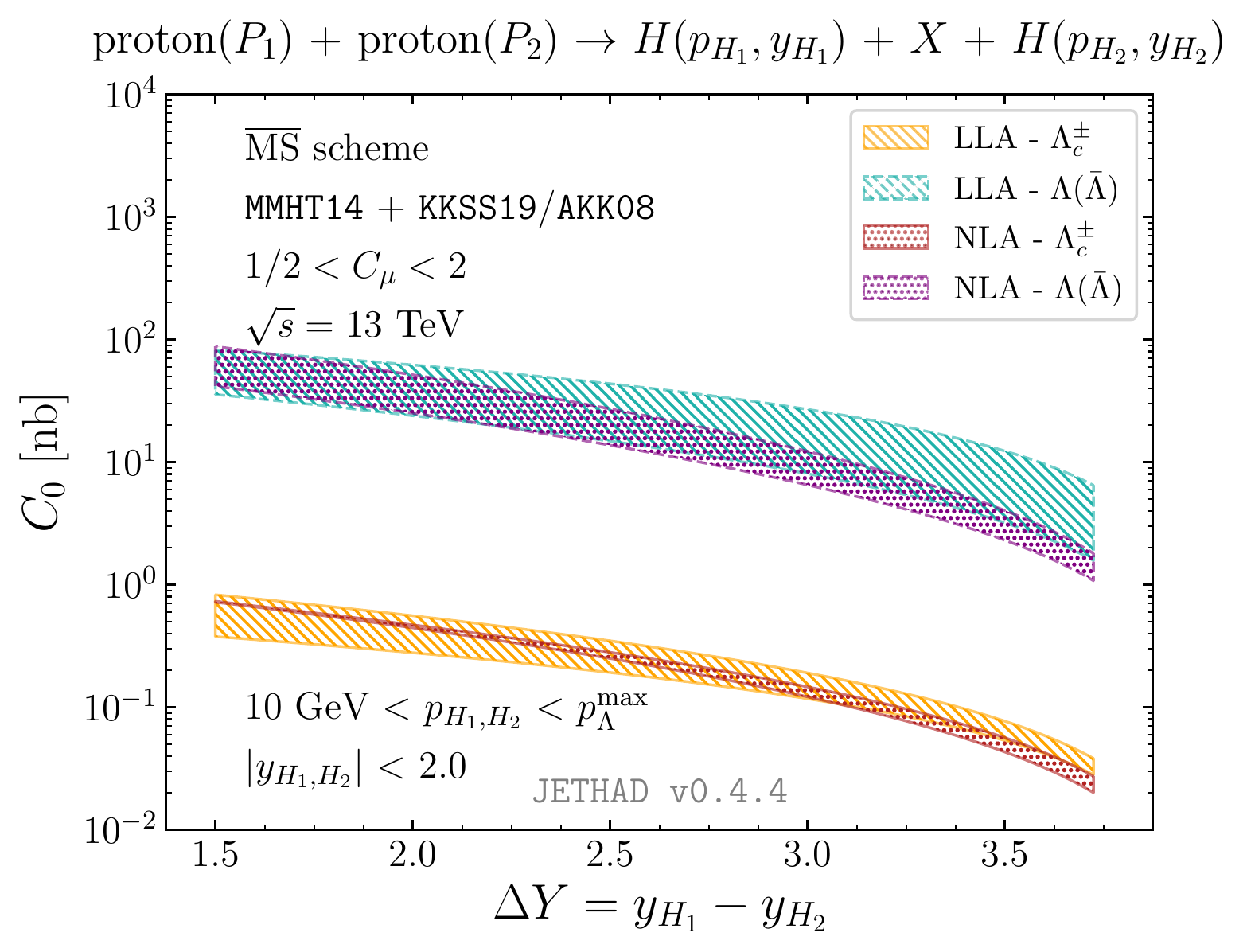}
   \includegraphics[scale=0.400,clip]{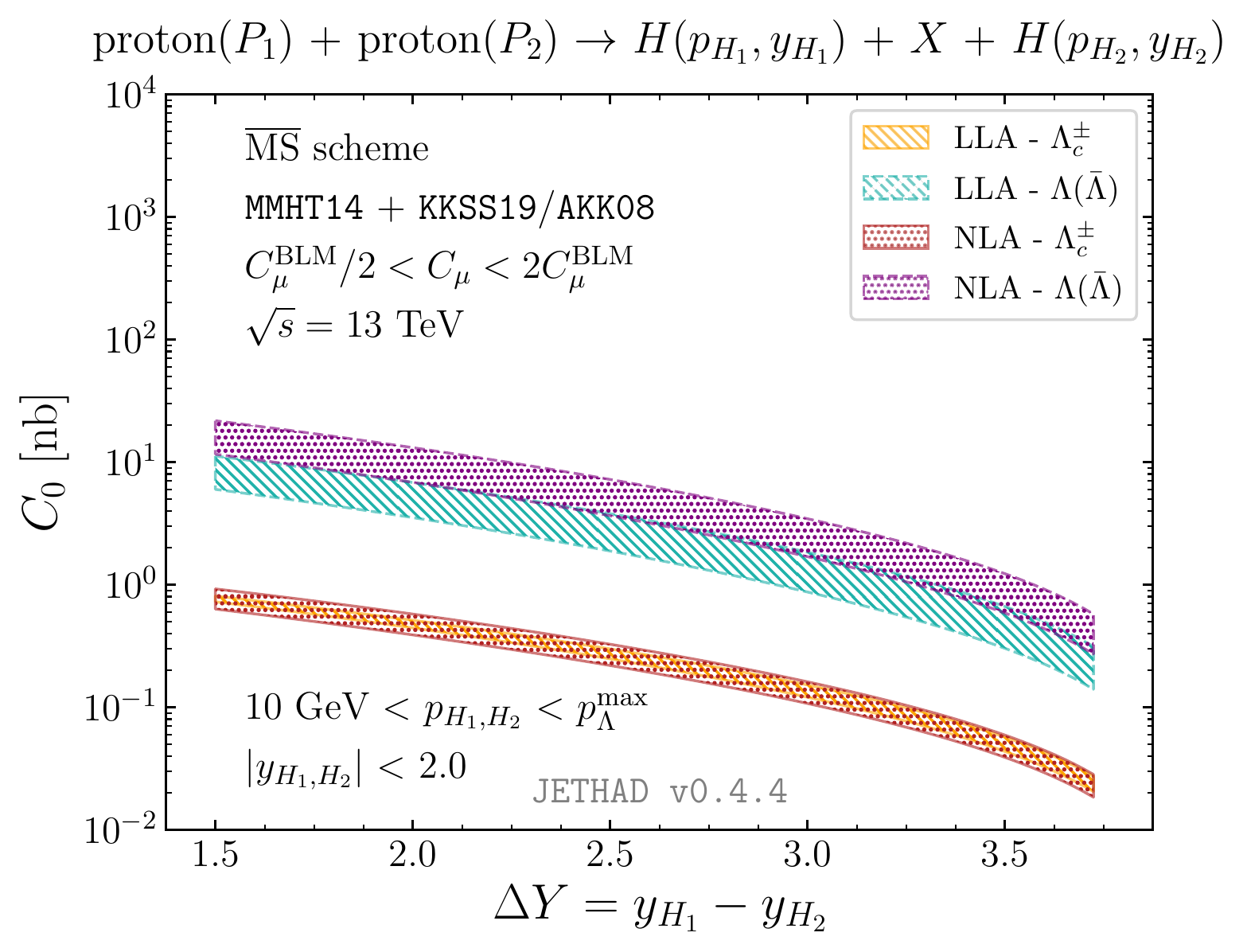}
   
   \includegraphics[scale=0.40,clip]{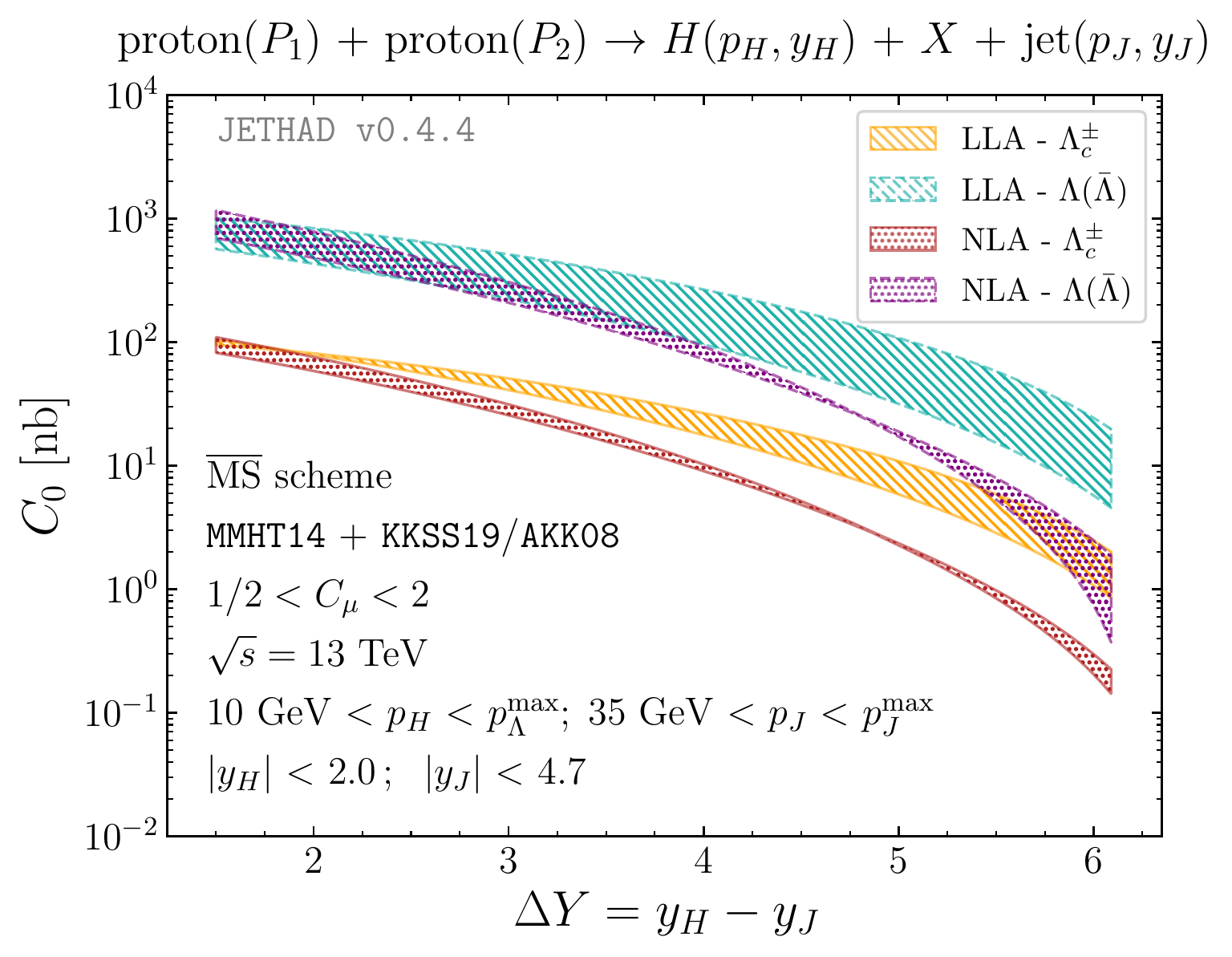}
   \includegraphics[scale=0.400,clip]{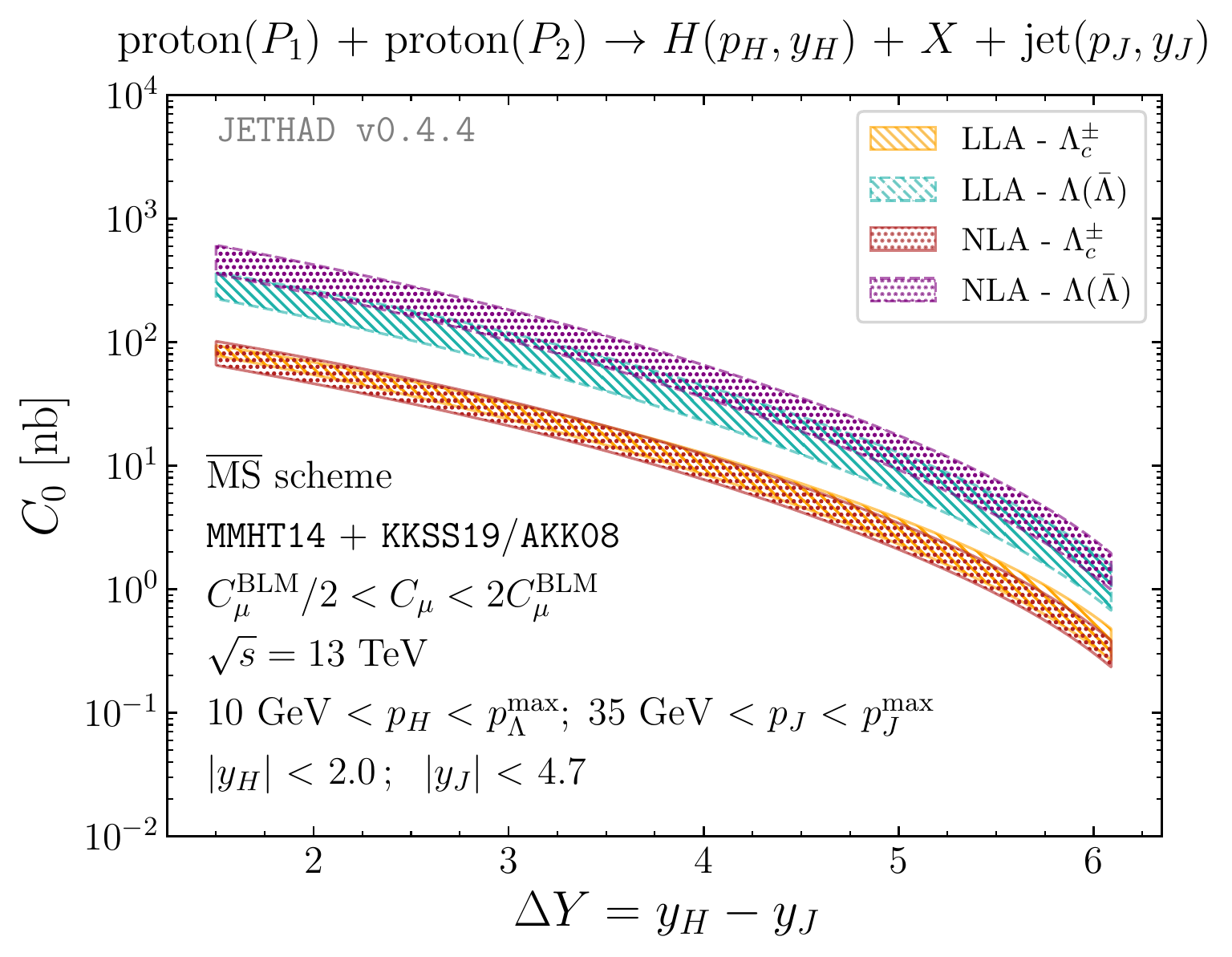}

\caption{Behavior of the $\varphi$-summed cross section, $C_0$, as a function of $\Delta Y$, in the double $\Lambda_c$ (upper) and in the $\Lambda_c$ plus jet channel (lower), at natural scales (left) and after BLM optimization (right), and for $\sqrt{s} = 13$ TeV. Error bands provide with the combined uncertainty coming from scale variation and numerical integrations. Predictions for $\Lambda_c$ emissions are compared with configurations where  $\Lambda$ hyperons are detected.}

\label{fig:C0_Lambda}
\end{figure}

\begin{figure}
\centering

   \includegraphics[scale=0.40,clip]{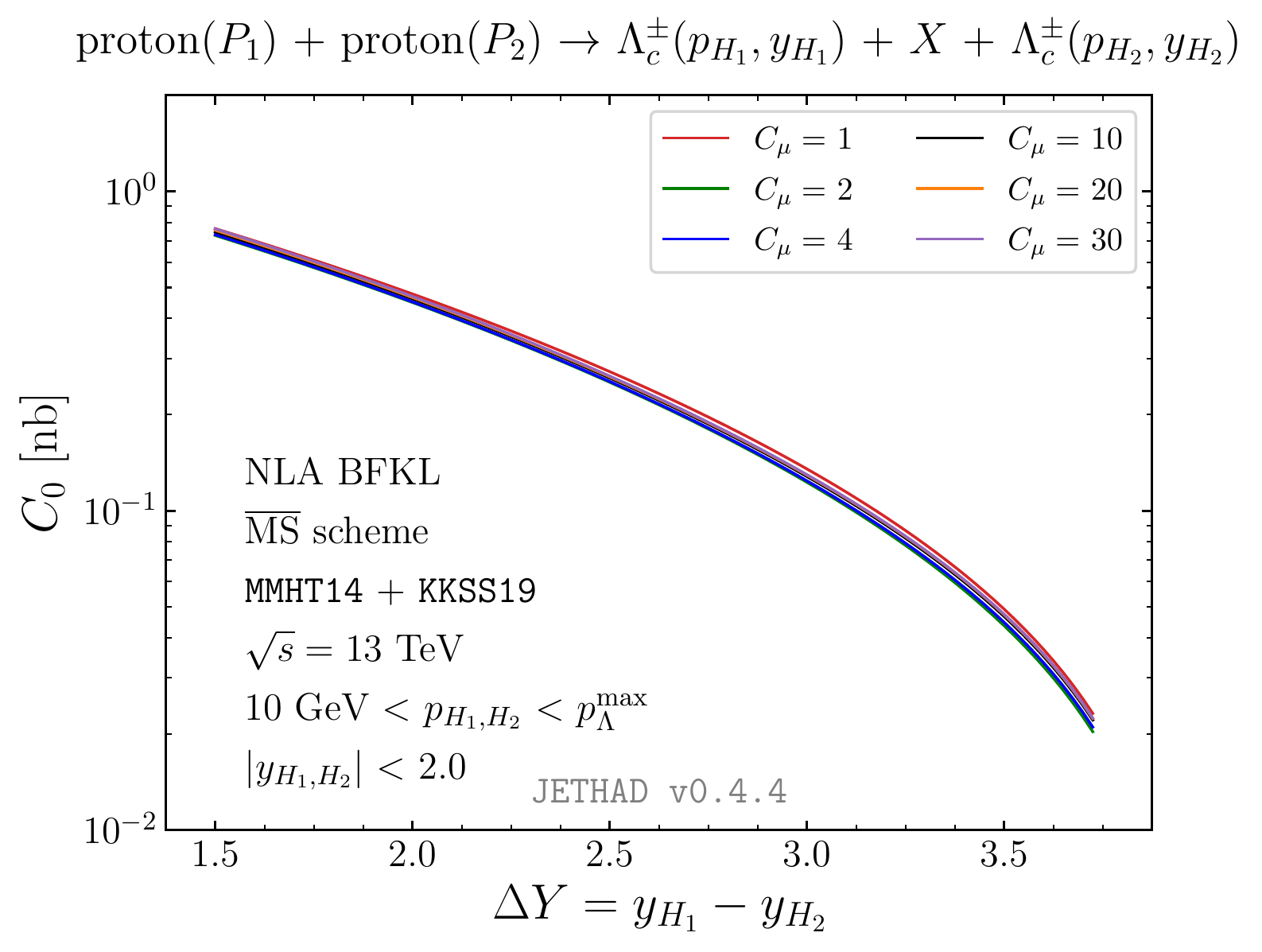}
   \includegraphics[scale=0.40,clip]{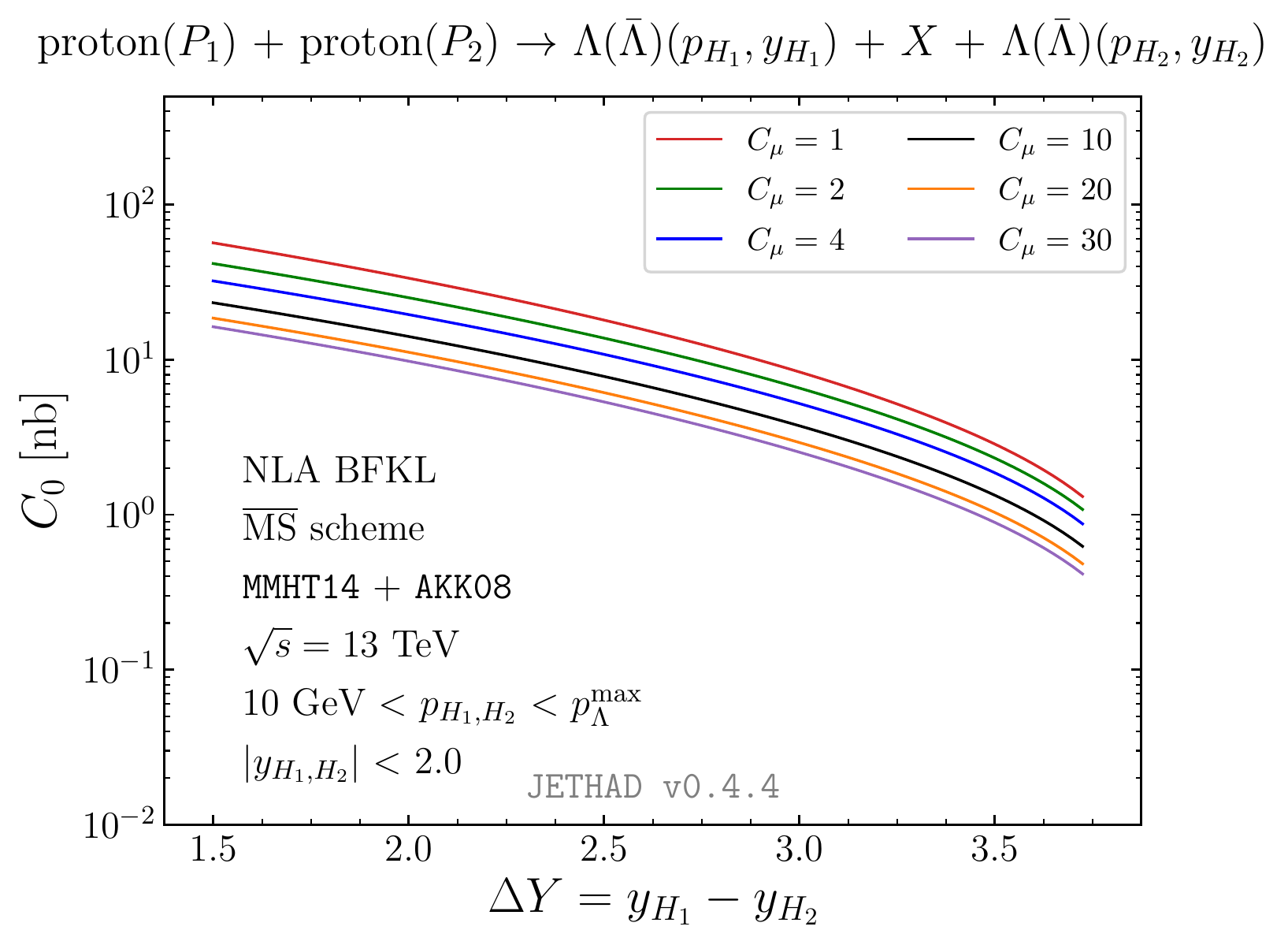}

\caption{Behavior of the $\varphi$-summed cross section, $C_0$, as a function of $\Delta Y$, in the dihadron production channel, and for $\sqrt{s} = 13$ TeV. 
A study on progressive energy-scale variation in the range $1 < C_{\mu} < 30$ is done for $\Lambda_c$ emissions (left) and for $\Lambda$ detections (right).}
\label{fig:C0_psv}
\end{figure}

To show the stabilization mechanism that occurs in the production of heavy species such as $\Lambda_c$, we will compare its cross section summed over azimuthal angles with the corresponding ones for lighter species. Key ingredients to build our distributions are the azimuthal coefficients integrated over rapidity and transverse momenta of the two tagged objects, and differential in the rapidity difference $\Delta Y \equiv y_1 - y_2$
\begin{equation}
 \label{Cn_int}
 C_n =
 \int_{y_1^{\rm min}}^{y_1^{\rm max}} d y_1
 \int_{y_2^{\rm min}}^{y_2^{\rm max}} d y_2
 \int_{p_1^{\rm min}}^{p_1^{\rm max}} d |\vec p_1|
 \int_{p_2^{\rm min}}^{p_2^{\rm max}} d |\vec p_2|
 \, \,
 \delta (\Delta Y - y_1 + y_2)
 \, \,
 {\cal C}_n\left(|\vec p_1|, |\vec p_2|, y_1, y_2 \right)
 \, .
\end{equation}
Here, we consider just $C_0$, whereas a more detailed analysis on the azimuthal correlation can be found in Ref.~\cite{Celiberto:2021dzy}. We impose LHC-typical kinematic cuts for both $\Lambda$-particles and jets, allowing the transverse momenta of the $\Lambda$ to range between $10$ GeV and $p_\Lambda^{{\rm{max}}} \simeq 21.5$ GeV and the jets one between $35$ GeV and $60$ GeV. As for the rapidities, we set $|y_\Lambda|<2.0$ and $|y_{J}|<4.7$. We fix the center-of-mass energy at $\sqrt{s} = 13$ TeV. We perform our phenomenological studies by making use of the {\tt JETHAD} modular interface~\cite{Celiberto:2020wpk} under development at our Group. We depict the parton fragmentation into $\Lambda_c$ baryons in terms of the {\tt KKSS19} NLO FF set \cite{Kniehl:2020szu}, while lighter-hadron emissions ($\Lambda$ hyperons) are described in terms of {\tt AKK08} NLO FFs \cite{Albino:2008fy}.  
In upper panels of Fig.~\ref{fig:C0_Lambda} we show the $\Delta Y$-dependence of the $\varphi$-summed cross section, $C_0$, in the double $\Lambda_c$ channel, together with corresponding predictions for the detection of $\Lambda$ hyperons.
We note that NLA bands are almost nested (except for very large values of $\Delta Y$) inside LLA ones and they are generally narrower in the $\Lambda_c$ case. This is a clear effect of a (partially) reached stability of the high-energy series, for both hadron emissions. However, while predictions for hyperons lose almost one order of magnitude when passing from natural scales to the expanded BLM ones (from left to right panel), results for $\Lambda_c$ baryons are much more stable, the NLA band becoming even wider in the BLM case.
The stability is partially lost when a $\Lambda_c$ particle is accompanied by a jet, as shown in lower panels of Fig.~\ref{fig:C0_Lambda}. Here, LLA and NLA bands are almost disjoined at natural scales (left panel), while in the BLM case (right panel) they come closer to each other for hyperon plus jet, and almost entirely contained for $\Lambda_c$ plus jet. 
In Fig.~\ref{fig:C0_psv} we study $C_0$ for the double production of $\Lambda_c$ baryons (left) or $\Lambda$ hyperons (right) under a progressive variation of energy scales in a wider range that includes the typical BLM ones, $1 < C_\mu < 30$. $C_0$ exhibits a fair stability under progressive scale variation both in the $\Lambda_c$, while its sensitivity spans over almost one order of magnitude in the hyperon case.
These studies on $C_0$ clearly highlight how $\Lambda_c$ emissions allow for a stabilization of the resummed series, that cannot be obtained with lighter hadrons. Further studies in~\cite{Celiberto:2021dzy,Celiberto:2021fdp} have evidenced how the stability effect is due to the smooth- and non-decreasing with $\mu_F$ behavior of $\Lambda_c$ FF. We plan to extend our program on semi-hard phenomenology by considering inclusive production of heavy-light mesons and quarkonia at the LHC and at new-generation colliding facilities~\cite{AbdulKhalek:2021gbh,Arbuzov:2020cqg,Chapon:2020heu,Anchordoqui:2021ghd}. 

\vspace{-0.25cm}
\bibliographystyle{apsrev}
\bibliography{references}

\begin{thebibliography}{48}
\expandafter\ifx\csname natexlab\endcsname\relax\def\natexlab#1{#1}\fi
\expandafter\ifx\csname bibnamefont\endcsname\relax
  \def\bibnamefont#1{#1}\fi
\expandafter\ifx\csname bibfnamefont\endcsname\relax
  \def\bibfnamefont#1{#1}\fi
\expandafter\ifx\csname citenamefont\endcsname\relax
  \def\citenamefont#1{#1}\fi
\expandafter\ifx\csname url\endcsname\relax
  \def\url#1{\texttt{#1}}\fi
\expandafter\ifx\csname urlprefix\endcsname\relax\def\urlprefix{URL }\fi
\providecommand{\bibinfo}[2]{#2}
\providecommand{\eprint}[2][]{\url{#2}}

\bibitem[{\citenamefont{Fadin et~al.}(1975)\citenamefont{Fadin, Kuraev, and
  Lipatov}}]{Fadin:1975cb}
\bibinfo{author}{\bibfnamefont{V.~S.} \bibnamefont{Fadin}},
  \bibinfo{author}{\bibfnamefont{E.}~\bibnamefont{Kuraev}}, \bibnamefont{and}
  \bibinfo{author}{\bibfnamefont{L.}~\bibnamefont{Lipatov}},
  \bibinfo{journal}{Phys. Lett. B} \textbf{\bibinfo{volume}{60}},
  \bibinfo{pages}{50} (\bibinfo{year}{1975}).

\bibitem[{\citenamefont{Kuraev et~al.}(1976)\citenamefont{Kuraev, Lipatov, and
  Fadin}}]{Kuraev:1976ge}
\bibinfo{author}{\bibfnamefont{E.~A.} \bibnamefont{Kuraev}},
  \bibinfo{author}{\bibfnamefont{L.~N.} \bibnamefont{Lipatov}},
  \bibnamefont{and} \bibinfo{author}{\bibfnamefont{V.~S.} \bibnamefont{Fadin}},
  \bibinfo{journal}{Sov. Phys. JETP} \textbf{\bibinfo{volume}{44}},
  \bibinfo{pages}{443} (\bibinfo{year}{1976}).

\bibitem[{\citenamefont{Kuraev et~al.}(1977)\citenamefont{Kuraev, Lipatov, and
  Fadin}}]{Kuraev:1977fs}
\bibinfo{author}{\bibfnamefont{E.}~\bibnamefont{Kuraev}},
  \bibinfo{author}{\bibfnamefont{L.}~\bibnamefont{Lipatov}}, \bibnamefont{and}
  \bibinfo{author}{\bibfnamefont{V.~S.} \bibnamefont{Fadin}},
  \bibinfo{journal}{Sov.\ Phys.\ JETP} \textbf{\bibinfo{volume}{45}},
  \bibinfo{pages}{199} (\bibinfo{year}{1977}).

\bibitem[{\citenamefont{Balitsky and Lipatov}(1978)}]{Balitsky:1978ic}
\bibinfo{author}{\bibfnamefont{I.}~\bibnamefont{Balitsky}} \bibnamefont{and}
  \bibinfo{author}{\bibfnamefont{L.}~\bibnamefont{Lipatov}},
  \bibinfo{journal}{Sov.\ J.\ Nucl.\ Phys.} \textbf{\bibinfo{volume}{28}},
  \bibinfo{pages}{822} (\bibinfo{year}{1978}).

\bibitem[{\citenamefont{Mueller and Navelet}(1987)}]{Mueller:1986ey}
\bibinfo{author}{\bibfnamefont{A.~H.} \bibnamefont{Mueller}} \bibnamefont{and}
  \bibinfo{author}{\bibfnamefont{H.}~\bibnamefont{Navelet}},
  \bibinfo{journal}{Nucl. Phys. B} \textbf{\bibinfo{volume}{282}},
  \bibinfo{pages}{727} (\bibinfo{year}{1987}).

\bibitem[{\citenamefont{Colferai et~al.}(2010)\citenamefont{Colferai,
  Schwennsen, Szymanowski, and Wallon}}]{Colferai:2010wu}
\bibinfo{author}{\bibfnamefont{D.}~\bibnamefont{Colferai}},
  \bibinfo{author}{\bibfnamefont{F.}~\bibnamefont{Schwennsen}},
  \bibinfo{author}{\bibfnamefont{L.}~\bibnamefont{Szymanowski}},
  \bibnamefont{and} \bibinfo{author}{\bibfnamefont{S.}~\bibnamefont{Wallon}},
  \bibinfo{journal}{JHEP} \textbf{\bibinfo{volume}{12}}, \bibinfo{pages}{026}
  (\bibinfo{year}{2010}), \eprint{1002.1365}.

\bibitem[{\citenamefont{Caporale et~al.}(2013)\citenamefont{Caporale, Ivanov,
  Murdaca, and Papa}}]{Caporale:2012ih}
\bibinfo{author}{\bibfnamefont{F.}~\bibnamefont{Caporale}},
  \bibinfo{author}{\bibfnamefont{D.~{\relax Yu}.} \bibnamefont{Ivanov}},
  \bibinfo{author}{\bibfnamefont{B.}~\bibnamefont{Murdaca}}, \bibnamefont{and}
  \bibinfo{author}{\bibfnamefont{A.}~\bibnamefont{Papa}},
  \bibinfo{journal}{Nucl. Phys. B} \textbf{\bibinfo{volume}{877}},
  \bibinfo{pages}{73} (\bibinfo{year}{2013}), \eprint{1211.7225}.

\bibitem[{\citenamefont{Duclou\'e et~al.}(2013)\citenamefont{Duclou\'e,
  Szymanowski, and Wallon}}]{Ducloue:2013hia}
\bibinfo{author}{\bibfnamefont{B.}~\bibnamefont{Duclou\'e}},
  \bibinfo{author}{\bibfnamefont{L.}~\bibnamefont{Szymanowski}},
  \bibnamefont{and} \bibinfo{author}{\bibfnamefont{S.}~\bibnamefont{Wallon}},
  \bibinfo{journal}{JHEP} \textbf{\bibinfo{volume}{05}}, \bibinfo{pages}{096}
  (\bibinfo{year}{2013}), \eprint{1302.7012}.

\bibitem[{\citenamefont{Duclou\'e et~al.}(2014)\citenamefont{Duclou\'e,
  Szymanowski, and Wallon}}]{Ducloue:2013bva}
\bibinfo{author}{\bibfnamefont{B.}~\bibnamefont{Duclou\'e}},
  \bibinfo{author}{\bibfnamefont{L.}~\bibnamefont{Szymanowski}},
  \bibnamefont{and} \bibinfo{author}{\bibfnamefont{S.}~\bibnamefont{Wallon}},
  \bibinfo{journal}{Phys. Rev. Lett.} \textbf{\bibinfo{volume}{112}},
  \bibinfo{pages}{082003} (\bibinfo{year}{2014}), \eprint{1309.3229}.

\bibitem[{\citenamefont{Caporale et~al.}(2014)\citenamefont{Caporale, Ivanov,
  Murdaca, and Papa}}]{Caporale:2014gpa}
\bibinfo{author}{\bibfnamefont{F.}~\bibnamefont{Caporale}},
  \bibinfo{author}{\bibfnamefont{D.~{\relax Yu}.} \bibnamefont{Ivanov}},
  \bibinfo{author}{\bibfnamefont{B.}~\bibnamefont{Murdaca}}, \bibnamefont{and}
  \bibinfo{author}{\bibfnamefont{A.}~\bibnamefont{Papa}},
  \bibinfo{journal}{Eur. Phys. J. C} \textbf{\bibinfo{volume}{74}},
  \bibinfo{pages}{3084} (\bibinfo{year}{2014}), \bibinfo{note}{[Erratum:
  Eur.Phys.J.C 75, 535 (2015)]}, \eprint{1407.8431}.

\bibitem[{\citenamefont{Caporale et~al.}(2015)\citenamefont{Caporale, Ivanov,
  Murdaca, and Papa}}]{Caporale:2015uva}
\bibinfo{author}{\bibfnamefont{F.}~\bibnamefont{Caporale}},
  \bibinfo{author}{\bibfnamefont{D.~{\relax Yu}.} \bibnamefont{Ivanov}},
  \bibinfo{author}{\bibfnamefont{B.}~\bibnamefont{Murdaca}}, \bibnamefont{and}
  \bibinfo{author}{\bibfnamefont{A.}~\bibnamefont{Papa}},
  \bibinfo{journal}{Phys. Rev. D} \textbf{\bibinfo{volume}{91}},
  \bibinfo{pages}{114009} (\bibinfo{year}{2015}), \eprint{1504.06471}.

\bibitem[{\citenamefont{Celiberto
  et~al.}(2015{\natexlab{a}})\citenamefont{Celiberto, Ivanov, Murdaca, and
  Papa}}]{Celiberto:2015yba}
\bibinfo{author}{\bibfnamefont{F.~G.} \bibnamefont{Celiberto}},
  \bibinfo{author}{\bibfnamefont{D.~{\relax Yu}.} \bibnamefont{Ivanov}},
  \bibinfo{author}{\bibfnamefont{B.}~\bibnamefont{Murdaca}}, \bibnamefont{and}
  \bibinfo{author}{\bibfnamefont{A.}~\bibnamefont{Papa}},
  \bibinfo{journal}{Eur. Phys. J. C} \textbf{\bibinfo{volume}{75}},
  \bibinfo{pages}{292} (\bibinfo{year}{2015}{\natexlab{a}}),
  \eprint{1504.08233}.

\bibitem[{\citenamefont{Celiberto
  et~al.}(2015{\natexlab{b}})\citenamefont{Celiberto, Ivanov, Murdaca, and
  Papa}}]{Celiberto:2015mpa}
\bibinfo{author}{\bibfnamefont{F.~G.} \bibnamefont{Celiberto}},
  \bibinfo{author}{\bibfnamefont{D.~{\relax Yu}.} \bibnamefont{Ivanov}},
  \bibinfo{author}{\bibfnamefont{B.}~\bibnamefont{Murdaca}}, \bibnamefont{and}
  \bibinfo{author}{\bibfnamefont{A.}~\bibnamefont{Papa}},
  \bibinfo{journal}{Acta Phys. Polon. Supp.} \textbf{\bibinfo{volume}{8}},
  \bibinfo{pages}{935} (\bibinfo{year}{2015}{\natexlab{b}}),
  \eprint{1510.01626}.

\bibitem[{\citenamefont{Celiberto
  et~al.}(2016{\natexlab{a}})\citenamefont{Celiberto, Ivanov, Murdaca, and
  Papa}}]{Celiberto:2016ygs}
\bibinfo{author}{\bibfnamefont{F.~G.} \bibnamefont{Celiberto}},
  \bibinfo{author}{\bibfnamefont{D.~{\relax Yu}.} \bibnamefont{Ivanov}},
  \bibinfo{author}{\bibfnamefont{B.}~\bibnamefont{Murdaca}}, \bibnamefont{and}
  \bibinfo{author}{\bibfnamefont{A.}~\bibnamefont{Papa}},
  \bibinfo{journal}{Eur. Phys. J. C} \textbf{\bibinfo{volume}{76}},
  \bibinfo{pages}{224} (\bibinfo{year}{2016}{\natexlab{a}}),
  \eprint{1601.07847}.

\bibitem[{\citenamefont{Celiberto
  et~al.}(2016{\natexlab{b}})\citenamefont{Celiberto, Ivanov, Murdaca, and
  Papa}}]{Celiberto:2016vva}
\bibinfo{author}{\bibfnamefont{F.~G.} \bibnamefont{Celiberto}},
  \bibinfo{author}{\bibfnamefont{D.~{\relax Yu}.} \bibnamefont{Ivanov}},
  \bibinfo{author}{\bibfnamefont{B.}~\bibnamefont{Murdaca}}, \bibnamefont{and}
  \bibinfo{author}{\bibfnamefont{A.}~\bibnamefont{Papa}},
  \bibinfo{journal}{PoS} \textbf{\bibinfo{volume}{DIS2016}},
  \bibinfo{pages}{176} (\bibinfo{year}{2016}{\natexlab{b}}),
  \eprint{1606.08892}.

\bibitem[{\citenamefont{Caporale et~al.}(2018)\citenamefont{Caporale,
  Celiberto, Chachamis, Gordo~Gómez, and Sabio~Vera}}]{Caporale:2018qnm}
\bibinfo{author}{\bibfnamefont{F.}~\bibnamefont{Caporale}},
  \bibinfo{author}{\bibfnamefont{F.~G.} \bibnamefont{Celiberto}},
  \bibinfo{author}{\bibfnamefont{G.}~\bibnamefont{Chachamis}},
  \bibinfo{author}{\bibfnamefont{D.}~\bibnamefont{Gordo~Gómez}},
  \bibnamefont{and}
  \bibinfo{author}{\bibfnamefont{A.}~\bibnamefont{Sabio~Vera}},
  \bibinfo{journal}{Nucl. Phys. B} \textbf{\bibinfo{volume}{935}},
  \bibinfo{pages}{412} (\bibinfo{year}{2018}), \eprint{1806.06309}.

\bibitem[{\citenamefont{Celiberto
  et~al.}(2016{\natexlab{c}})\citenamefont{Celiberto, Ivanov, Murdaca, and
  Papa}}]{Celiberto:2016hae}
\bibinfo{author}{\bibfnamefont{F.~G.} \bibnamefont{Celiberto}},
  \bibinfo{author}{\bibfnamefont{D.~{\relax Yu}.} \bibnamefont{Ivanov}},
  \bibinfo{author}{\bibfnamefont{B.}~\bibnamefont{Murdaca}}, \bibnamefont{and}
  \bibinfo{author}{\bibfnamefont{A.}~\bibnamefont{Papa}},
  \bibinfo{journal}{Phys. Rev. D} \textbf{\bibinfo{volume}{94}},
  \bibinfo{pages}{034013} (\bibinfo{year}{2016}{\natexlab{c}}),
  \eprint{1604.08013}.

\bibitem[{\citenamefont{Celiberto
  et~al.}(2017{\natexlab{a}})\citenamefont{Celiberto, Ivanov, Murdaca, and
  Papa}}]{Celiberto:2016zgb}
\bibinfo{author}{\bibfnamefont{F.~G.} \bibnamefont{Celiberto}},
  \bibinfo{author}{\bibfnamefont{D.~{\relax Yu}.} \bibnamefont{Ivanov}},
  \bibinfo{author}{\bibfnamefont{B.}~\bibnamefont{Murdaca}}, \bibnamefont{and}
  \bibinfo{author}{\bibfnamefont{A.}~\bibnamefont{Papa}}, \bibinfo{journal}{AIP
  Conf. Proc.} \textbf{\bibinfo{volume}{1819}}, \bibinfo{pages}{060005}
  (\bibinfo{year}{2017}{\natexlab{a}}), \eprint{1611.04811}.

\bibitem[{\citenamefont{Celiberto
  et~al.}(2017{\natexlab{b}})\citenamefont{Celiberto, Ivanov, Murdaca, and
  Papa}}]{Celiberto:2017ptm}
\bibinfo{author}{\bibfnamefont{F.~G.} \bibnamefont{Celiberto}},
  \bibinfo{author}{\bibfnamefont{D.~{\relax Yu}.} \bibnamefont{Ivanov}},
  \bibinfo{author}{\bibfnamefont{B.}~\bibnamefont{Murdaca}}, \bibnamefont{and}
  \bibinfo{author}{\bibfnamefont{A.}~\bibnamefont{Papa}},
  \bibinfo{journal}{Eur. Phys. J. C} \textbf{\bibinfo{volume}{77}},
  \bibinfo{pages}{382} (\bibinfo{year}{2017}{\natexlab{b}}),
  \eprint{1701.05077}.

\bibitem[{\citenamefont{Celiberto
  et~al.}(2017{\natexlab{c}})\citenamefont{Celiberto, Ivanov, Murdaca, and
  Papa}}]{Celiberto:2017ydk}
\bibinfo{author}{\bibfnamefont{F.~G.} \bibnamefont{Celiberto}},
  \bibinfo{author}{\bibfnamefont{D.~{\relax Yu}.} \bibnamefont{Ivanov}},
  \bibinfo{author}{\bibfnamefont{B.}~\bibnamefont{Murdaca}}, \bibnamefont{and}
  \bibinfo{author}{\bibfnamefont{A.}~\bibnamefont{Papa}}, in
  \emph{\bibinfo{booktitle}{{17th conference on Elastic and Diffractive
  Scattering}}} (\bibinfo{year}{2017}{\natexlab{c}}), \eprint{1709.04758}.

\bibitem[{\citenamefont{Bolognino
  et~al.}(2018{\natexlab{a}})\citenamefont{Bolognino, Celiberto, Ivanov,
  Mohammed, and Papa}}]{Bolognino:2018oth}
\bibinfo{author}{\bibfnamefont{A.~D.} \bibnamefont{Bolognino}},
  \bibinfo{author}{\bibfnamefont{F.~G.} \bibnamefont{Celiberto}},
  \bibinfo{author}{\bibfnamefont{D.~{\relax Yu}.} \bibnamefont{Ivanov}},
  \bibinfo{author}{\bibfnamefont{M.~M.} \bibnamefont{Mohammed}},
  \bibnamefont{and} \bibinfo{author}{\bibfnamefont{A.}~\bibnamefont{Papa}},
  \bibinfo{journal}{Eur. Phys. J. C} \textbf{\bibinfo{volume}{78}},
  \bibinfo{pages}{772} (\bibinfo{year}{2018}{\natexlab{a}}),
  \eprint{1808.05483}.

\bibitem[{\citenamefont{Bolognino
  et~al.}(2019{\natexlab{a}})\citenamefont{Bolognino, Celiberto, Ivanov,
  Mohammed, and Papa}}]{Bolognino:2019yqj}
\bibinfo{author}{\bibfnamefont{A.~D.} \bibnamefont{Bolognino}},
  \bibinfo{author}{\bibfnamefont{F.~G.} \bibnamefont{Celiberto}},
  \bibinfo{author}{\bibfnamefont{D.~{\relax Yu}.} \bibnamefont{Ivanov}},
  \bibinfo{author}{\bibfnamefont{M.~M.} \bibnamefont{Mohammed}},
  \bibnamefont{and} \bibinfo{author}{\bibfnamefont{A.}~\bibnamefont{Papa}},
  \bibinfo{journal}{Acta Phys. Polon. Supp.} \textbf{\bibinfo{volume}{12}},
  \bibinfo{pages}{773} (\bibinfo{year}{2019}{\natexlab{a}}),
  \eprint{1902.04511}.

\bibitem[{\citenamefont{Bolognino
  et~al.}(2019{\natexlab{b}})\citenamefont{Bolognino, Celiberto, Ivanov,
  Mohammed, and Papa}}]{Bolognino:2019cac}
\bibinfo{author}{\bibfnamefont{A.~D.} \bibnamefont{Bolognino}},
  \bibinfo{author}{\bibfnamefont{F.~G.} \bibnamefont{Celiberto}},
  \bibinfo{author}{\bibfnamefont{D.~{\relax Yu}.} \bibnamefont{Ivanov}},
  \bibinfo{author}{\bibfnamefont{M.~M.~A.} \bibnamefont{Mohammed}},
  \bibnamefont{and} \bibinfo{author}{\bibfnamefont{A.}~\bibnamefont{Papa}},
  \bibinfo{journal}{PoS} \textbf{\bibinfo{volume}{DIS2019}},
  \bibinfo{pages}{049} (\bibinfo{year}{2019}{\natexlab{b}}),
  \eprint{1906.11800}.

\bibitem[{\citenamefont{Celiberto et~al.}(2020)\citenamefont{Celiberto, Ivanov,
  and Papa}}]{Celiberto:2020rxb}
\bibinfo{author}{\bibfnamefont{F.~G.} \bibnamefont{Celiberto}},
  \bibinfo{author}{\bibfnamefont{D.~{\relax Yu}.} \bibnamefont{Ivanov}},
  \bibnamefont{and} \bibinfo{author}{\bibfnamefont{A.}~\bibnamefont{Papa}},
  \bibinfo{journal}{Phys. Rev. D} \textbf{\bibinfo{volume}{102}},
  \bibinfo{pages}{094019} (\bibinfo{year}{2020}), \eprint{2008.10513}.

\bibitem[{\citenamefont{Caporale
  et~al.}(2016{\natexlab{a}})\citenamefont{Caporale, Chachamis, Murdaca, and
  Sabio~Vera}}]{Caporale:2015vya}
\bibinfo{author}{\bibfnamefont{F.}~\bibnamefont{Caporale}},
  \bibinfo{author}{\bibfnamefont{G.}~\bibnamefont{Chachamis}},
  \bibinfo{author}{\bibfnamefont{B.}~\bibnamefont{Murdaca}}, \bibnamefont{and}
  \bibinfo{author}{\bibfnamefont{A.}~\bibnamefont{Sabio~Vera}},
  \bibinfo{journal}{Phys. Rev. Lett.} \textbf{\bibinfo{volume}{116}},
  \bibinfo{pages}{012001} (\bibinfo{year}{2016}{\natexlab{a}}),
  \eprint{1508.07711}.

\bibitem[{\citenamefont{Caporale
  et~al.}(2016{\natexlab{b}})\citenamefont{Caporale, Celiberto, Chachamis, and
  Sabio~Vera}}]{Caporale:2015int}
\bibinfo{author}{\bibfnamefont{F.}~\bibnamefont{Caporale}},
  \bibinfo{author}{\bibfnamefont{F.~G.} \bibnamefont{Celiberto}},
  \bibinfo{author}{\bibfnamefont{G.}~\bibnamefont{Chachamis}},
  \bibnamefont{and}
  \bibinfo{author}{\bibfnamefont{A.}~\bibnamefont{Sabio~Vera}},
  \bibinfo{journal}{Eur. Phys. J. C} \textbf{\bibinfo{volume}{76}},
  \bibinfo{pages}{165} (\bibinfo{year}{2016}{\natexlab{b}}),
  \eprint{1512.03364}.

\bibitem[{\citenamefont{Caporale
  et~al.}(2016{\natexlab{c}})\citenamefont{Caporale, Celiberto, Chachamis,
  Gordo~Gómez, and Sabio~Vera}}]{Caporale:2016soq}
\bibinfo{author}{\bibfnamefont{F.}~\bibnamefont{Caporale}},
  \bibinfo{author}{\bibfnamefont{F.~G.} \bibnamefont{Celiberto}},
  \bibinfo{author}{\bibfnamefont{G.}~\bibnamefont{Chachamis}},
  \bibinfo{author}{\bibfnamefont{D.}~\bibnamefont{Gordo~Gómez}},
  \bibnamefont{and}
  \bibinfo{author}{\bibfnamefont{A.}~\bibnamefont{Sabio~Vera}},
  \bibinfo{journal}{Nucl. Phys. B} \textbf{\bibinfo{volume}{910}},
  \bibinfo{pages}{374} (\bibinfo{year}{2016}{\natexlab{c}}),
  \eprint{1603.07785}.

\bibitem[{\citenamefont{Caporale
  et~al.}(2017{\natexlab{a}})\citenamefont{Caporale, Celiberto, Chachamis,
  Gordo~Gómez, and Sabio~Vera}}]{Caporale:2016xku}
\bibinfo{author}{\bibfnamefont{F.}~\bibnamefont{Caporale}},
  \bibinfo{author}{\bibfnamefont{F.~G.} \bibnamefont{Celiberto}},
  \bibinfo{author}{\bibfnamefont{G.}~\bibnamefont{Chachamis}},
  \bibinfo{author}{\bibfnamefont{D.}~\bibnamefont{Gordo~Gómez}},
  \bibnamefont{and}
  \bibinfo{author}{\bibfnamefont{A.}~\bibnamefont{Sabio~Vera}},
  \bibinfo{journal}{Eur. Phys. J. C} \textbf{\bibinfo{volume}{77}},
  \bibinfo{pages}{5} (\bibinfo{year}{2017}{\natexlab{a}}), \eprint{1606.00574}.

\bibitem[{\citenamefont{Caporale
  et~al.}(2017{\natexlab{b}})\citenamefont{Caporale, Celiberto, Chachamis,
  Gordo~Gómez, and Sabio~Vera}}]{Caporale:2016zkc}
\bibinfo{author}{\bibfnamefont{F.}~\bibnamefont{Caporale}},
  \bibinfo{author}{\bibfnamefont{F.~G.} \bibnamefont{Celiberto}},
  \bibinfo{author}{\bibfnamefont{G.}~\bibnamefont{Chachamis}},
  \bibinfo{author}{\bibfnamefont{D.}~\bibnamefont{Gordo~Gómez}},
  \bibnamefont{and}
  \bibinfo{author}{\bibfnamefont{A.}~\bibnamefont{Sabio~Vera}},
  \bibinfo{journal}{Phys. Rev. D} \textbf{\bibinfo{volume}{95}},
  \bibinfo{pages}{074007} (\bibinfo{year}{2017}{\natexlab{b}}),
  \eprint{1612.05428}.

\bibitem[{\citenamefont{Boussarie et~al.}(2018)\citenamefont{Boussarie,
  Duclou\'e, Szymanowski, and Wallon}}]{Boussarie:2017oae}
\bibinfo{author}{\bibfnamefont{R.}~\bibnamefont{Boussarie}},
  \bibinfo{author}{\bibfnamefont{B.}~\bibnamefont{Duclou\'e}},
  \bibinfo{author}{\bibfnamefont{L.}~\bibnamefont{Szymanowski}},
  \bibnamefont{and} \bibinfo{author}{\bibfnamefont{S.}~\bibnamefont{Wallon}},
  \bibinfo{journal}{Phys. Rev. D} \textbf{\bibinfo{volume}{97}},
  \bibinfo{pages}{014008} (\bibinfo{year}{2018}), \eprint{1709.01380}.

\bibitem[{\citenamefont{Golec-Biernat et~al.}(2018)\citenamefont{Golec-Biernat,
  Motyka, and Stebel}}]{Golec-Biernat:2018kem}
\bibinfo{author}{\bibfnamefont{K.}~\bibnamefont{Golec-Biernat}},
  \bibinfo{author}{\bibfnamefont{L.}~\bibnamefont{Motyka}}, \bibnamefont{and}
  \bibinfo{author}{\bibfnamefont{T.}~\bibnamefont{Stebel}},
  \bibinfo{journal}{JHEP} \textbf{\bibinfo{volume}{12}}, \bibinfo{pages}{091}
  (\bibinfo{year}{2018}), \eprint{1811.04361}.

\bibitem[{\citenamefont{Celiberto
  et~al.}(2021{\natexlab{a}})\citenamefont{Celiberto, Ivanov, Mohammed, and
  Papa}}]{Celiberto:2020tmb}
\bibinfo{author}{\bibfnamefont{F.~G.} \bibnamefont{Celiberto}},
  \bibinfo{author}{\bibfnamefont{D.~{\relax Yu}.} \bibnamefont{Ivanov}},
  \bibinfo{author}{\bibfnamefont{M.~M.~A.} \bibnamefont{Mohammed}},
  \bibnamefont{and} \bibinfo{author}{\bibfnamefont{A.}~\bibnamefont{Papa}},
  \bibinfo{journal}{Eur. Phys. J. C} \textbf{\bibinfo{volume}{81}},
  \bibinfo{pages}{293} (\bibinfo{year}{2021}{\natexlab{a}}),
  \eprint{2008.00501}.

\bibitem[{\citenamefont{Celiberto
  et~al.}(2018{\natexlab{a}})\citenamefont{Celiberto, Ivanov, Murdaca, and
  Papa}}]{Celiberto:2017nyx}
\bibinfo{author}{\bibfnamefont{F.~G.} \bibnamefont{Celiberto}},
  \bibinfo{author}{\bibfnamefont{D.~{\relax Yu}.} \bibnamefont{Ivanov}},
  \bibinfo{author}{\bibfnamefont{B.}~\bibnamefont{Murdaca}}, \bibnamefont{and}
  \bibinfo{author}{\bibfnamefont{A.}~\bibnamefont{Papa}},
  \bibinfo{journal}{Phys. Lett. B} \textbf{\bibinfo{volume}{777}},
  \bibinfo{pages}{141} (\bibinfo{year}{2018}{\natexlab{a}}),
  \eprint{1709.10032}.

\bibitem[{\citenamefont{Bolognino
  et~al.}(2019{\natexlab{c}})\citenamefont{Bolognino, Celiberto, Fucilla,
  Ivanov, Murdaca, and Papa}}]{Bolognino:2019ouc}
\bibinfo{author}{\bibfnamefont{A.~D.} \bibnamefont{Bolognino}},
  \bibinfo{author}{\bibfnamefont{F.~G.} \bibnamefont{Celiberto}},
  \bibinfo{author}{\bibfnamefont{M.}~\bibnamefont{Fucilla}},
  \bibinfo{author}{\bibfnamefont{D.~{\relax Yu}.} \bibnamefont{Ivanov}},
  \bibinfo{author}{\bibfnamefont{B.}~\bibnamefont{Murdaca}}, \bibnamefont{and}
  \bibinfo{author}{\bibfnamefont{A.}~\bibnamefont{Papa}},
  \bibinfo{journal}{PoS} \textbf{\bibinfo{volume}{DIS2019}},
  \bibinfo{pages}{067} (\bibinfo{year}{2019}{\natexlab{c}}),
  \eprint{1906.05940}.

\bibitem[{\citenamefont{Bolognino
  et~al.}(2019{\natexlab{d}})\citenamefont{Bolognino, Celiberto, Fucilla,
  Ivanov, and Papa}}]{Bolognino:2019yls}
\bibinfo{author}{\bibfnamefont{A.~D.} \bibnamefont{Bolognino}},
  \bibinfo{author}{\bibfnamefont{F.~G.} \bibnamefont{Celiberto}},
  \bibinfo{author}{\bibfnamefont{M.}~\bibnamefont{Fucilla}},
  \bibinfo{author}{\bibfnamefont{D.~{\relax Yu}.} \bibnamefont{Ivanov}},
  \bibnamefont{and} \bibinfo{author}{\bibfnamefont{A.}~\bibnamefont{Papa}},
  \bibinfo{journal}{Eur. Phys. J. C} \textbf{\bibinfo{volume}{79}},
  \bibinfo{pages}{939} (\bibinfo{year}{2019}{\natexlab{d}}),
  \eprint{1909.03068}.

\bibitem[{\citenamefont{Bolognino
  et~al.}(2021{\natexlab{a}})\citenamefont{Bolognino, Celiberto, Fucilla,
  Ivanov, and Papa}}]{Bolognino:2021mrc}
\bibinfo{author}{\bibfnamefont{A.~D.} \bibnamefont{Bolognino}},
  \bibinfo{author}{\bibfnamefont{F.~G.} \bibnamefont{Celiberto}},
  \bibinfo{author}{\bibfnamefont{M.}~\bibnamefont{Fucilla}},
  \bibinfo{author}{\bibfnamefont{D.~{\relax Yu}.} \bibnamefont{Ivanov}},
  \bibnamefont{and} \bibinfo{author}{\bibfnamefont{A.}~\bibnamefont{Papa}},
  \bibinfo{journal}{Phys. Rev. D} \textbf{\bibinfo{volume}{103}},
  \bibinfo{pages}{094004} (\bibinfo{year}{2021}{\natexlab{a}}),
  \eprint{2103.07396}.

\bibitem[{\citenamefont{Bolognino
  et~al.}(2018{\natexlab{b}})\citenamefont{Bolognino, Celiberto, Ivanov, and
  Papa}}]{Bolognino:2018rhb}
\bibinfo{author}{\bibfnamefont{A.~D.} \bibnamefont{Bolognino}},
  \bibinfo{author}{\bibfnamefont{F.~G.} \bibnamefont{Celiberto}},
  \bibinfo{author}{\bibfnamefont{D.~{\relax Yu}.} \bibnamefont{Ivanov}},
  \bibnamefont{and} \bibinfo{author}{\bibfnamefont{A.}~\bibnamefont{Papa}},
  \bibinfo{journal}{Eur. Phys. J.} \textbf{\bibinfo{volume}{C78}},
  \bibinfo{pages}{1023} (\bibinfo{year}{2018}{\natexlab{b}}),
  \eprint{1808.02395}.

\bibitem[{\citenamefont{Celiberto
  et~al.}(2018{\natexlab{b}})\citenamefont{Celiberto, Gordo~Gomez, and
  Sabio~Vera}}]{Celiberto:2018muu}
\bibinfo{author}{\bibfnamefont{F.~G.} \bibnamefont{Celiberto}},
  \bibinfo{author}{\bibfnamefont{D.}~\bibnamefont{Gordo~Gomez}},
  \bibnamefont{and}
  \bibinfo{author}{\bibfnamefont{A.}~\bibnamefont{Sabio~Vera}},
  \bibinfo{journal}{Phys. Lett.} \textbf{\bibinfo{volume}{B786}},
  \bibinfo{pages}{201} (\bibinfo{year}{2018}{\natexlab{b}}),
  \eprint{1808.09511}.

\bibitem[{\citenamefont{Bolognino
  et~al.}(2021{\natexlab{b}})\citenamefont{Bolognino, Celiberto, Ivanov, Papa,
  Sch\"afer, and Szczurek}}]{Bolognino:2021niq}
\bibinfo{author}{\bibfnamefont{A.~D.} \bibnamefont{Bolognino}},
  \bibinfo{author}{\bibfnamefont{F.~G.} \bibnamefont{Celiberto}},
  \bibinfo{author}{\bibfnamefont{D.~{\relax Yu}.} \bibnamefont{Ivanov}},
  \bibinfo{author}{\bibfnamefont{A.}~\bibnamefont{Papa}},
  \bibinfo{author}{\bibfnamefont{W.}~\bibnamefont{Sch\"afer}},
  \bibnamefont{and} \bibinfo{author}{\bibfnamefont{A.}~\bibnamefont{Szczurek}},
  \bibinfo{journal}{Eur. Phys. J. C} \textbf{\bibinfo{volume}{81}},
  \bibinfo{pages}{846} (\bibinfo{year}{2021}{\natexlab{b}}),
  \eprint{2107.13415}.

\bibitem[{\citenamefont{Celiberto
  et~al.}(2021{\natexlab{b}})\citenamefont{Celiberto, Fucilla, Ivanov, and
  Papa}}]{Celiberto:2021dzy}
\bibinfo{author}{\bibfnamefont{F.~G.} \bibnamefont{Celiberto}},
  \bibinfo{author}{\bibfnamefont{M.}~\bibnamefont{Fucilla}},
  \bibinfo{author}{\bibfnamefont{D.~{\relax Yu}.} \bibnamefont{Ivanov}},
  \bibnamefont{and} \bibinfo{author}{\bibfnamefont{A.}~\bibnamefont{Papa}},
  \bibinfo{journal}{Eur. Phys. J. C} \textbf{\bibinfo{volume}{81}},
  \bibinfo{pages}{780} (\bibinfo{year}{2021}{\natexlab{b}}),
  \eprint{2105.06432}.

\bibitem[{\citenamefont{Celiberto}(2021)}]{Celiberto:2020wpk}
\bibinfo{author}{\bibfnamefont{F.~G.} \bibnamefont{Celiberto}},
  \bibinfo{journal}{Eur. Phys. J. C} \textbf{\bibinfo{volume}{81}},
  \bibinfo{pages}{691} (\bibinfo{year}{2021}), \eprint{2008.07378}.

\bibitem[{\citenamefont{Kniehl et~al.}(2020)\citenamefont{Kniehl, Kramer,
  Schienbein, and Spiesberger}}]{Kniehl:2020szu}
\bibinfo{author}{\bibfnamefont{B.~A.} \bibnamefont{Kniehl}},
  \bibinfo{author}{\bibfnamefont{G.}~\bibnamefont{Kramer}},
  \bibinfo{author}{\bibfnamefont{I.}~\bibnamefont{Schienbein}},
  \bibnamefont{and}
  \bibinfo{author}{\bibfnamefont{H.}~\bibnamefont{Spiesberger}},
  \bibinfo{journal}{Phys. Rev. D} \textbf{\bibinfo{volume}{101}},
  \bibinfo{pages}{114021} (\bibinfo{year}{2020}), \eprint{2004.04213}.

\bibitem[{\citenamefont{Albino et~al.}(2008)\citenamefont{Albino, Kniehl, and
  Kramer}}]{Albino:2008fy}
\bibinfo{author}{\bibfnamefont{S.}~\bibnamefont{Albino}},
  \bibinfo{author}{\bibfnamefont{B.~A.} \bibnamefont{Kniehl}},
  \bibnamefont{and} \bibinfo{author}{\bibfnamefont{G.}~\bibnamefont{Kramer}},
  \bibinfo{journal}{Nucl. Phys. B} \textbf{\bibinfo{volume}{803}},
  \bibinfo{pages}{42} (\bibinfo{year}{2008}), \eprint{0803.2768}.

\bibitem[{\citenamefont{Celiberto
  et~al.}(2021{\natexlab{c}})\citenamefont{Celiberto, Fucilla, Ivanov,
  Mohammed, and Papa}}]{Celiberto:2021fdp}
\bibinfo{author}{\bibfnamefont{F.~G.} \bibnamefont{Celiberto}},
  \bibinfo{author}{\bibfnamefont{M.}~\bibnamefont{Fucilla}},
  \bibinfo{author}{\bibfnamefont{D.~{\relax Yu}.} \bibnamefont{Ivanov}},
  \bibinfo{author}{\bibfnamefont{M.~M.~A.} \bibnamefont{Mohammed}},
  \bibnamefont{and} \bibinfo{author}{\bibfnamefont{A.}~\bibnamefont{Papa}}
  (\bibinfo{year}{2021}{\natexlab{c}}), \eprint{2109.11875}.

\bibitem[{\citenamefont{Abdul~Khalek et~al.}(2021)}]{AbdulKhalek:2021gbh}
\bibinfo{author}{\bibfnamefont{R.}~\bibnamefont{Abdul~Khalek}}
  \bibnamefont{et~al.} (\bibinfo{year}{2021}), \eprint{2103.05419}.

\bibitem[{\citenamefont{Arbuzov et~al.}(2021)}]{Arbuzov:2020cqg}
\bibinfo{author}{\bibfnamefont{A.}~\bibnamefont{Arbuzov}} \bibnamefont{et~al.},
  \bibinfo{journal}{Prog. Part. Nucl. Phys.} \textbf{\bibinfo{volume}{119}},
  \bibinfo{pages}{103858} (\bibinfo{year}{2021}), \eprint{2011.15005}.

\bibitem[{\citenamefont{Chapon et~al.}(2021)}]{Chapon:2020heu}
\bibinfo{author}{\bibfnamefont{E.}~\bibnamefont{Chapon}} \bibnamefont{et~al.},
  \bibinfo{journal}{Prog. Part. Nucl. Phys. (in press)}
  (\bibinfo{year}{2021}), \eprint{2012.14161}.

\bibitem[{\citenamefont{Anchordoqui et~al.}(2021)}]{Anchordoqui:2021ghd}
\bibinfo{author}{\bibfnamefont{L.~A.} \bibnamefont{Anchordoqui}}
  \bibnamefont{et~al.} (\bibinfo{year}{2021}), \eprint{2109.10905}.

\end{thebibliography}

\end{document}